\documentstyle{mn}

\title{Colour changes in quasar light curves}
\author[M. R. S. Hawkins]
       {M. R. S. Hawkins \\
        University of Edinburgh, Royal Observatory, Blackford Hill,
        Edinburgh EH9 3HJ, Scotland, UK}
\date{Accepted ??.
      Received ??;
      in original form ??}

\pagerange{\pageref{firstpage}--\pageref{lastpage}}
\pubyear{2002}

\begin{document}
\label{lastpage}

\maketitle

\label{firstpage}

\begin{abstract}

The study of quasar variability has long been seen as a way to
understanding the structure of the central engine of active galactic
nuclei, and as a means of verifying the morphology of the standard
model.  Much work has already been done on the statistical properties
of light curves monitored in one colour, and it is now possible to
use these observations to test predictions of theoretical models.
The addition of a second colour can add enormously to the power of
such tests, and put tight constraints on the nature of the variations.
In this paper a yearly monitoring programme of several hundred quasars
in blue and red passbands covering 21 years is presented.  The
statistics of colour changes are examined for a 15 year period of
homogeneous data with Fourier power spectrum analysis,
in a form suitable for testing against theoretical predictions.  The
results of the Fourier analysis show that there is more power in blue
light on all timescales than in the red.  Examination of the light
curves shows several different modes of colour change.  However, if
allowance is made for the effects of the underlying host galaxy, the
variations become close to achromatie.  There are however structural
differences between red and blue light curves which cannot be accounted
for in this way, and various modes of variability including disc
instability and microlensing are examined to provide explanations for
these features.

\end{abstract}


\section{Introduction}

\begin{figure*}
\begin{picture} (560,400) (0,0)
\includegraphics{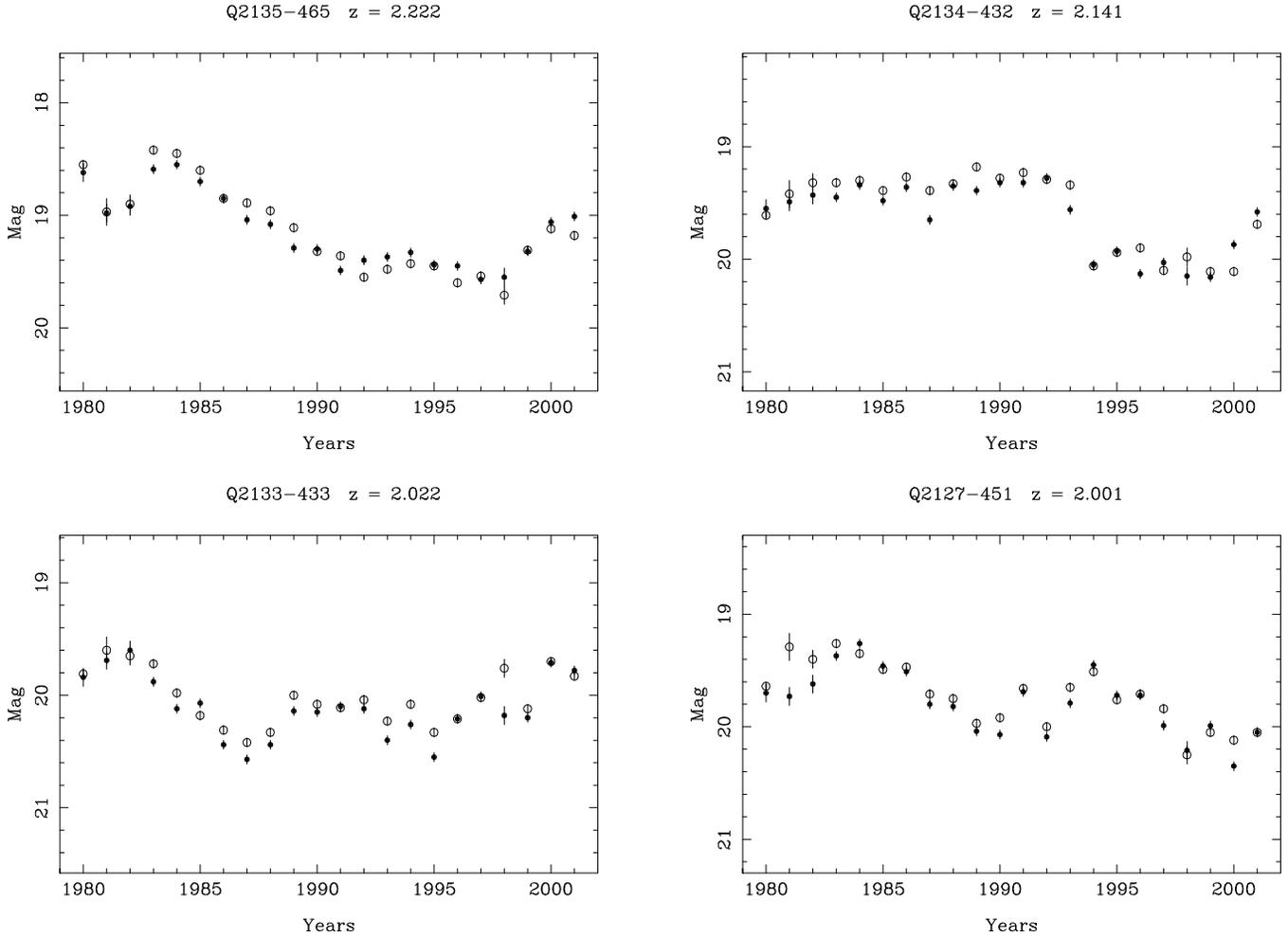}
\end{picture}
\caption{Light curves in $B_{J}$ (filled circles) and $R$ (open
 circles) for a selection of quasars which show no systematic colour
 change with brightness.
 \label{fig:fig1}}
\end{figure*}

It has long been recognised that the variability of active galactic
nuclei (AGN) has the potential to probe the structure and dynamics
of the central engine and put tight constraints on any unified
model.  For low luminosity AGN (Seyfert galaxies) extensive monitoring
of short term photometric and spectroscopic variations \cite{p99}
has made a major impact on the understanding of the structure of the
central regions.  Much effort has also been put into the optical
monitoring of luminous AGN (quasars) \cite{h94,c96,h96}, but with
less concrete results.  The main problems have been the long time
scales over which quasars appear to vary making uniformly sampled
light curves for a large number of objects very hard to obtain,
and the lack of theoretical predictions against which to test models
of quasar variability.  A first step towards rectifying this was
made possible by the publication of predicted structure functions for
light curves in accretion disc and starburst models of quasar
variability \cite{k98}.  The predictions were made for one colour
passband, and were suitable for comparing with structure functions
for samples of quasar light curves.  What was needed was a sample
of several hundred quasars covering a wide range of redshifts and
luminosities, regularly monitored over a period of decades.  Such
a sample has gradually been built up from plates taken with the
UK 1.2m Schmidt telescope in the ESO/SERC field 287, details of which
are given in the following section.  Light curves from these data,
together with theoretical predictions for the structure function of
variations, have been used in a recent paper
\cite{h02} to distinguish between various models of AGN variability.
An unexpectedly complicated picture emerged in which Seyfert galaxy
variations were best explained by an accretion disk model, while
quasar variability was closer to the predictions for microlensing.

A new dimension can be added to the existing analysis by the addition
of a second passband to the monitoring data.  Different models
of variability show very different behaviour in the way colour
changes manifest themselves in the light curves.  So far, quantitative
predictions suitable for testing against observations have not been
published for any models, but even so there are generic expectations
for colour effects in the best studied models.  For example, in
the starburst model for AGN variability \cite{a94}, the variations
are attributed to successions of supernova explosions.  In this model
one can confidently predict that AGN will become bluer as they get
brighter, just as supernovae do.

Observational work on colour changes in AGN is rather sparse.  The
long term monitoring programme of the Seyfert galaxy NGC 5548 has
shown strong colour changes in the ultraviolet from {\it IUE}
observations \cite{c91}.  This is illustrated in Fig. 3 of Hawkins
\& Taylor (1997) which shows the data uniformly sampled and converted
to a magnitude scale.  The situation with regard to quasars is less
clear.  Hawkins (1996) concluded that light variations in most
quasars were close to being achromatic, on the basis of an analysis
of correlations between yearly changes in $B_{J}$ and $R$ magnitudes,
but Cristiani et al. (1997) claimed to find a difference in the
amplitude of structure functions for $B$ and $R$ band light curves.
A similar small effect is seen in Fourier power spectra \cite{h01},
but in these data the author concludes that the difference is not
significant.

In this paper the intention is to improve on these earlier results
by carrying out a comprehensive analysis of colour changes in quasar
light curves.  This will provide a powerful discriminant between
quasar models as they become sufficiently robust to make testable
predictions.

\section{The quasar monitoring programme}

\begin{figure*}
\begin{picture} (560,600) (0,0)
\includegraphics{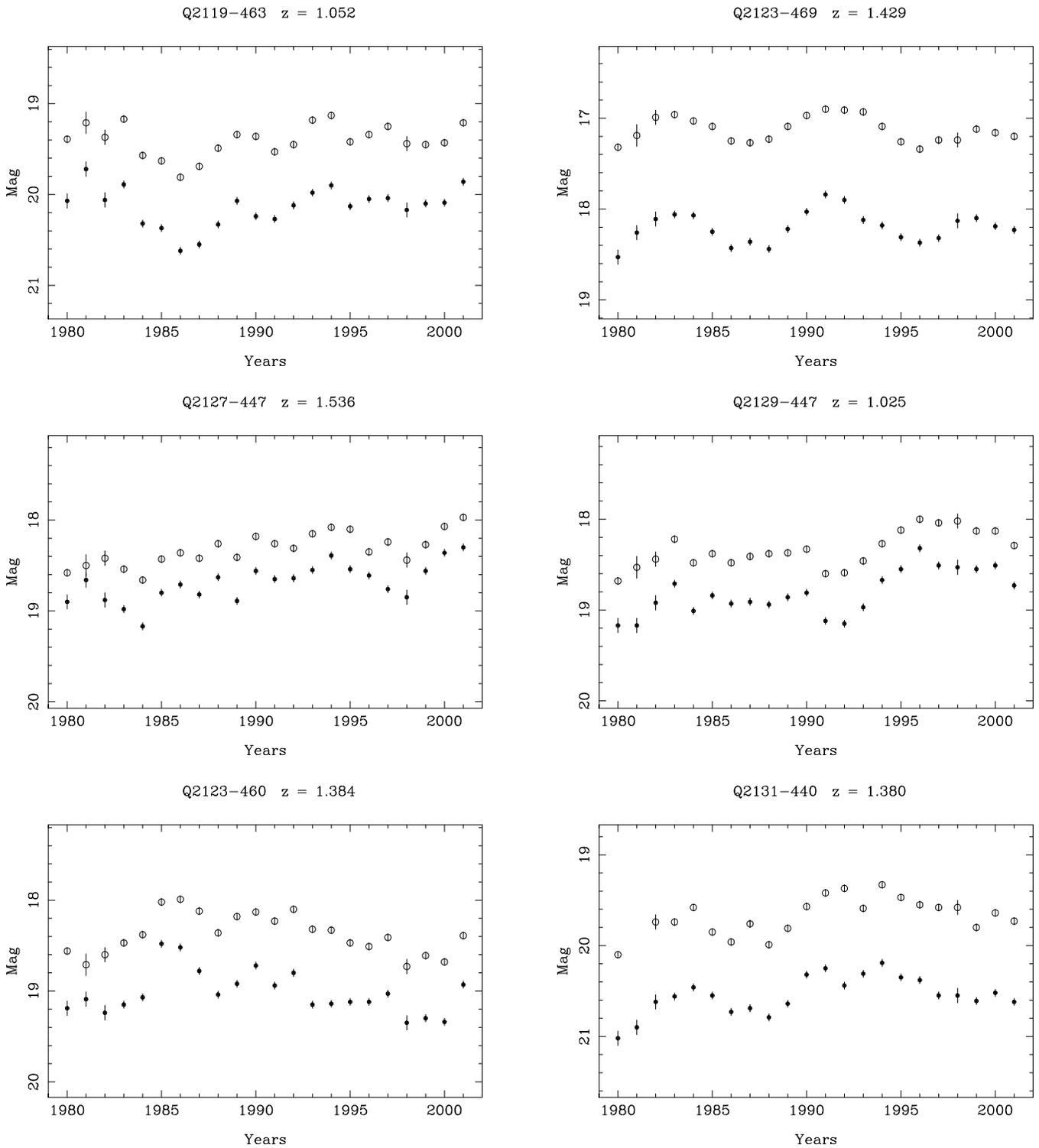}
\end{picture}
\caption{Light curves in $B_{J}$ (filled circles) and $R$ (open
 circles) for a typical selection of quasars.  The variation is
 largely achromatic.
 \label{fig:fig2}}
\end{figure*}

The light curves in this study are from a long term quasar monitoring
programme based on photographic plates taken with the UK Schmidt
Telescope of the ESO/SERC field 287.  The field centre is at 21h 28m,
-45$^{\circ}$ (1950), and some 350 sky limited plates and films have
been taken over a variety of timescales from hours to 26 years, and in
several optical passbands.  Of particular relevamce to the present
programme is a yearly sequence of observations in $B_{J}$ (IIIa-J
emulsion with a GG395 filter) and $R$ (IIIa-F or 4415 emulsion with a
RG630 filter) passbands covering 21 years from 1980 to 2001.  This
unbroken sequence normally included 4 sky limited exposures in both
$B_{J}$ and $R$ each year, but in a few years it was only possible to
obtain one plate in one or both of the passbands.  Details of all the
plates used in this paper are given in Table 1 of the Appendix.  The
early plates (up to 1992) were measured with the COSMOS measuring
machine, and after that with SuperCOSMOS at the Royal Observatory,
Edinburgh.  The survey area was confined to a square area of 19
deg$^{2}$ in the centre of the field to avoid problems with vignetting
and other off-axis effects.  Approximately 180,000 objects were
detected in this area to a limit of $B_{J} \approx 21.5$.  The $R$
plates went to a limit of $R \approx 20.5$.

The plates were paired up to give around 100 photometric measures in
each of $B_{J}$ and $R$ for every object on the plate where a detection
was made.  Because variability was the main interest for these
data, the measures were first reduced to a common photometric zero
point using local standards.  These transformed measures were then
calibrated with a CCD standards \cite{h96}, giving a typical relative
photometric error of 0.08 mag for each plate.  For most years where
4 plates were available this error reduced to 0.04 mag.  A more
detailed discussion of these errors is given by Hawkins (1996) and
references therein.  The absolute photometric errors across the field
are somewhat larger than this, but are hard to pin down precisely
without a grid of standards covering the whole field.  They are however
not significant for the present investigation where relative change in
brightness is being analysed.

The quasars for the sample were selected by a number of methods
including ultra-violet excess, variability, objective prism, and red
drop-out.  Of the estimated 1800 quasars in the field, some 1200 now
have redshifts in the range $0.1 < z < 4.5$.  From this parent
population a number of complete samples have been constructed according
to well-defined selection criteria \cite{h00}.  The light curves now
span 26 years in the $B_{J}$ passband, and 21 years in both $B_{J}$
and $R$.  Nearly all quasars discovered by colour techniques
have varied by more than 0.35 mag, the threshold for detection by
variability.  The statistics of this are discussed in more detail by
Hawkins (2000).  The quasar light curves in this survey appear to be
most useful on a timescale of 5 years or more.  Short term variations
were studied by taking 16 exposures over six months in 1984.  Typical
light curves are illustrated by Hawkins (1996), and analysis shows that
very few of the quasars in the sample vary by more than 0.1 mag over
this period, This is too close to the photometric errors on the
measurements to be useful for statistical analysis.  In fact the median
amplitude for quasar variation is about 0.6 mag, and most quasars in
the sample take a few years to achieve this \cite{h00}.

\section{Colour changes in quasar light curves}

\subsection{Qualitative aspects of the light curves}

It was stated by Hawkins (1996) that the variation of most quasars
appears to be nearly achromatic, and indeed it is not difficult to
find examples of quasar light curves which show no measurable colour
change even with large changes in amplitude.  Four such cases are
shown in Fig.~\ref{fig:fig1}, and it will be seen that although there
is some scatter between blue and red magnitudes in individual years,
there is no sytematic change in colour with brightness.  At the other
extreme, there are quasars which show light curves of very different
character in blue and red passbands \cite{h98}.

The long term (several years) variation of most quasars is near to
being achromatic, but close examination of the light curves shows small
departures from this simple picture.  Fig.~\ref{fig:fig2} shows six
typical quasar light curves which illustrate this.  It will be seen
that although there is no systematic colour change between maximum and
minimum brightness, sharp features in the blue light curves tend to be
slightly `smeared out' in the red.  Also, there are occasional small
isolated fluctuations on a timescale of a year in either passband which
are not mirrored in the other colour.  One type of variation which
is rarely seen in the quasar population is the strictly chromatic
variability observed for NGC 5548 \cite{c91}.

\subsection{Fourier power spectra of the light curves}

\begin{figure*}
\begin{picture} (560,200) (0,0)
\includegraphics{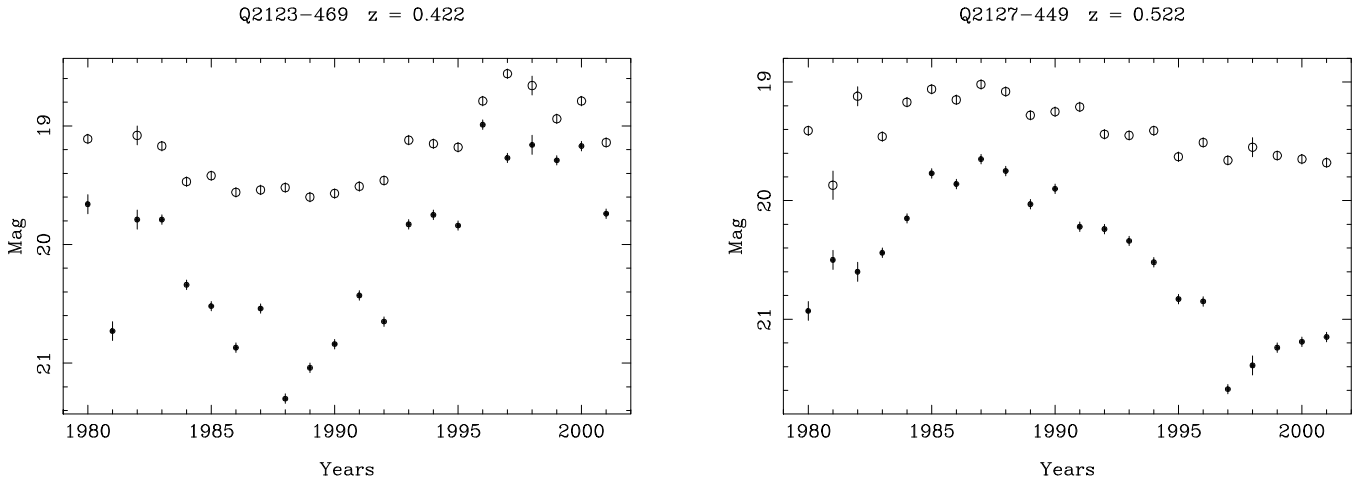}
\end{picture}
\caption{Light curves in $B_{J}$ (filled circles) and $R$ (open
 circles) for two low luminosity AGN which show the effects of the
 underlying galaxy in their light curves.
 \label{fig:fig3}}
\end{figure*}

\begin{figure*}
\begin{picture} (560,400) (0,0)
\includegraphics{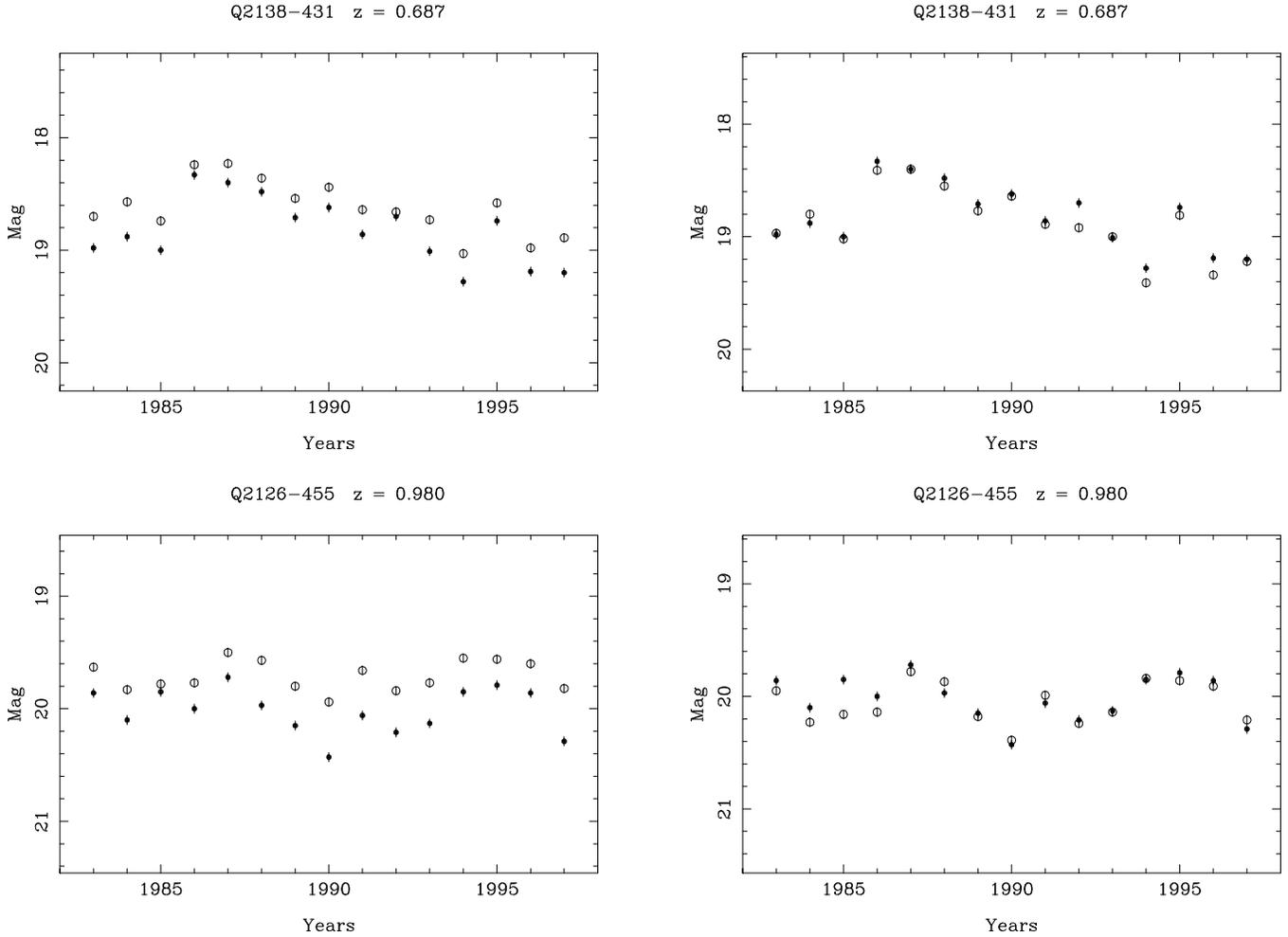}
\end{picture}
\caption{Light curves in $B_{J}$ (filled circles) and $R$ (open
 circles) for two AGN.  The left hand panels show the light curves
 as measured, with evidence for chromatic variation.  In the right hand
 panels the contribution from the underlying galaxy has been
 subtracted, and the variation is close to being achromatic.
 \label{fig:fig4}}
\end{figure*}

\begin{figure*}
\begin{picture} (560,280) (0,280)
\includegraphics{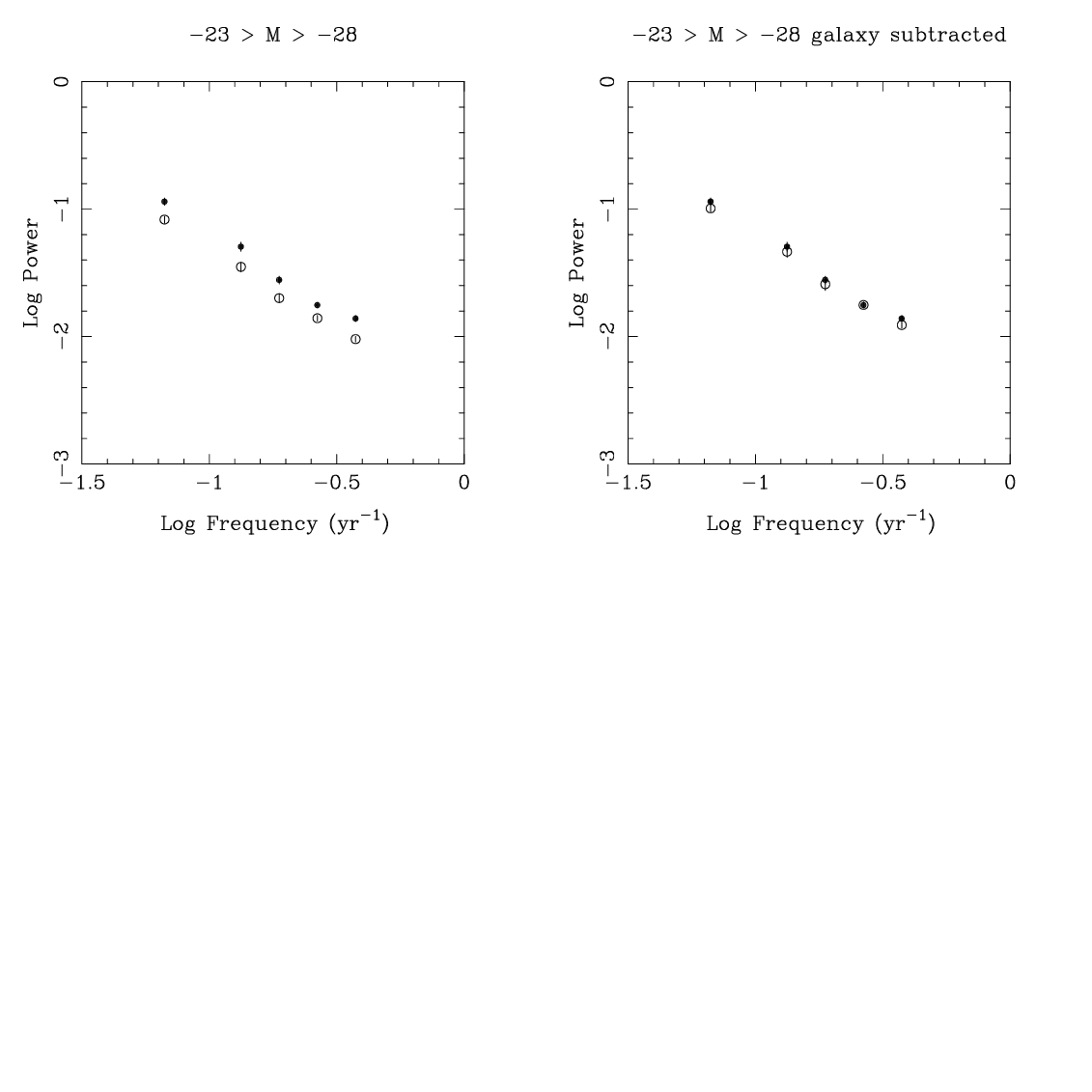}
\end{picture}
\caption{Fourier power spectra for quasar light curves in $B_{J}$
 (filled circles) and $R$ (open circles).  The left hand panel shows
 results for the observed light curves, and the right hand panel for
 the same light curves after subtracting a contribution for the
 underlying galaxy of absolute magnitude $M_{R} = -22.6$.
 \label{fig:fig5}}
\end{figure*}

Most statistical analysis of AGN light curves until recently was
carried out by calculating the structure function \cite{c96,h94} or
auto-correlation function \cite{h96}.  The idea was to characterise
the timescale or amplitude by reference to the shape of these
functions, but the early results did not
shed much light on these or any other parameters.  There were
several problems to overcome.  Perhaps the most important one concerns
the timescale covered by the observations.  Inspection of the light
curves in Figs 1 and 2 strongly suggests power on a timescale of
decades or more, and recent work has confirmed this \cite{h01}.
To properly measure the timescale of variation it is clear that a
run of data covering many tens of years is ideally needed.  Although
this ideal has yet to be achieved, the survey on which the present
analysis is based goes a long way to addressing the problem.  Another
problem with early datasets was the lack of homogeneous, regular
monitoring.  Without this, time series analysis becomes difficult, with
spurious features and aliassing dominating the results.  It is also
important to have a large sample of light curves covering a large span
of AGN redshift and luminosity to enable subsamples to be analysed and
compared.  Again, this survey marks a major improvement on earlier
work.

Another area of difficulty concerns the procedures which have been used
to analyse light curves.  The irregular pattern of observations in the
monitoring programmes has led to the choice of structure or
auto-correlations functions as analytical tools, which are relatively
insensitive to aliassing problems from unevenly spaced data.  However,
these functions have a number of drawbacks, including correlation
between the points, difficulty in characterising the errors, and
ambiguity of interpretation.  Fourier power spectra, although
containing essentially the same information for an infinite run of
data, provide a number of advantages for finite datasets.  The
frequency values are much less affected by neighbouring points, the
errors are relatively easy to calculate, and the interpretation is much
more straightforward than for other functions.  The main reason that
they have not been used in the past is that to be useful a long run of
evenly sampled homogeneous data is essential to avoid insurnountable
aliassing problems.  The data described in section 2 now meet these
requirements, and so we shall use Fourier power spectrum analysis to
investigate the chromatic properties of AGN light curves in this paper.

We define the Fourier power spectrum $P(j)$ for each light curve, in
the observers frame, as

\[ P(j) = \left\{\sum_{n=0}^{N-1} \cos\left(\frac{2\pi jn}{N}\right)
 m(n)\right\}^{2}+\left\{\sum_{n=0}^{N-1} \sin\left(\frac{2\pi jn}{N}
 \right) m(n)\right\}^{2}\]
where N is the number of epochs, and the frequency j runs from 0 to
N-1. The power spectrum of a sample of light curves is defined as the
mean of the individual power spectra P(j).

Fig.~\ref{fig:fig5} shows the combined Fourier power spectrum for a
sample of 302 blue and red
passband light curves for the period 1983-1987.  During this period
there were four plates of similar quality in both colours taken within
a few weeks of each other every year.  COSMOS measures of the four
yearly plates in each colour were averaged with equal weight to give
a uniformly sampled and homogeneous run of data.  The power spectrum
was evaluated for each light curve, and these power spectra were then
averaged as described above.  The errors  were calculated in two ways.
Firstly, the sample was split into various subsamples, and the
dispersion in the resulting power spectra used to estimate the errors.
Secondly, a sample of stars from the same set of measurements as the
quasars was analysed in the same way, and the dispersion in the
frequency measures used to estimate the error on each point.  The
errors obtained by these two methods were not significantly different,
and an average value was used.  The left hand panel shows data for the
observed light curves, and it will be seen that in both $B_{J}$ and $R$
passbands the shape of the power spectrum is close to a power law, and
although the two slopes are similar there is significantly more power
in $B_{J}$.  This implies a larger ampltude of variation in blue light,
and hence that the AGN become bluer as they brighten. 

\subsection{Effect of an underlying host galaxy}

One obvious possible reason for a change in colour as AGN vary is
the presence of an underlying galaxy of comparable brightness to the
nucleus, but of a redder colour.  When the nucleus is at its brightest,
the additional light from the underlying galaxy will have little effect
on either the blue or red light curves.  However, as the nucleus dims
the light of the quasar will become progressively more dominated by the
light of the underlying galaxy.  For a red galaxy this will happen
first  in the red, and the red light curve may even bottom out at the
$R$ magnitude of the underlying galaxy.  In blue light the quasar will
continue to fade, as the nuclear light continues to dominate the faint
$B$ flux of the galaxy.  Plausible examples of this process may be
seen in Fig.~\ref{fig:fig3} which shows the light curves for two low
luminosity AGN ($M_{B} \sim -22$).  This pattern of behaviour is never
seen in more luminous AGN (quasars) such as in Fig.~\ref{fig:fig1}.

The possibility that colour changes in light curves are caused by the
effects of an underlying galaxy can be tested by attempting to
correct for such an effect.  The first step is to estimate the mean
magnitude of the quasar host galaxies and its dispersion.  This can be
done for the quasars in the sample used here by looking for light
curves similar to those in Fig.~\ref{fig:fig3} which bottom out in the
red, but continue falling in blue.  A sample of 17 was chosen by eye
from the available quasar light curves, on the basis that they showed
clear signs of the variation in $R$ reaching a minimum when the $B_{J}$
light continued to fade.  The mean luminosity of the underlying
galaxies measured in this way was found to be $M_{R} = -22.6$ with a
standard deviation $\sigma = 0.4$.

The small dispersion in the flux of host galaxies provides a plausible
basis for attempting to correct for the colour distortion of the
light curves by subtracting the mean value from
each point in the $R$ band light curves.  The effect of doing this is
shown in Fig.~\ref{fig:fig4}.  Light curves for two quasars showing
evidence for colour changes with luminosity are shown in the two left
hand panels.  The colour change is such that the amplitude in $B_{J}$
is larger than in $R$.  The same light curves are shown in the two right
hand panels, but with the estimated mean flux of the underlying galaxy
subtracted.  It will be seen that apart from one or two isolated short
term fluctuations the variation is now close to being achromatic in
both light curves.

\begin{figure*}
\begin{picture} (560,560) (0,0)
\includegraphics{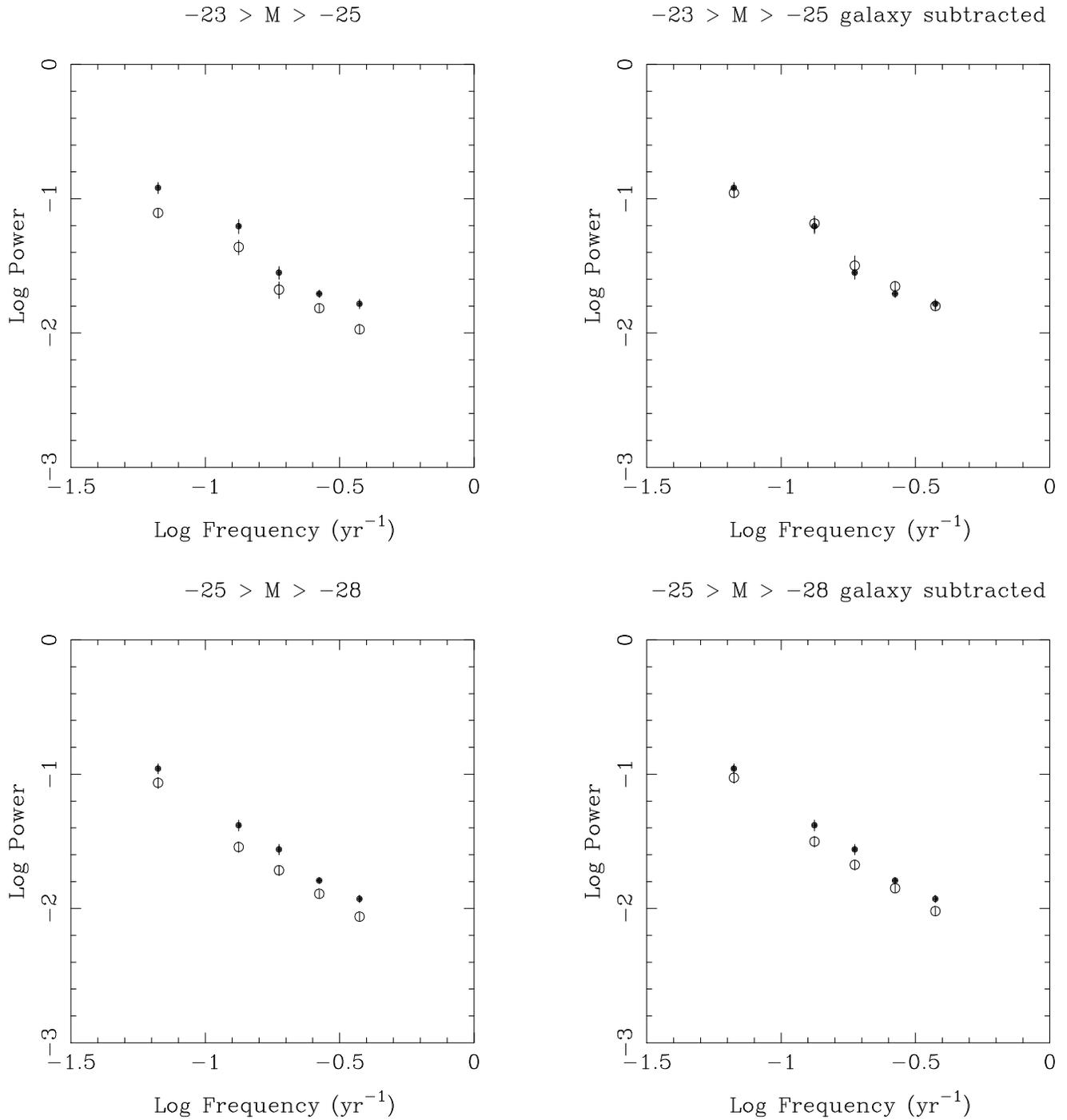}
\end{picture}
\caption{Fourier power spectra for quasar light curves in $B_{J}$
 (filled circles) and $R$ (open circles).  The left hand panels show
 results for the observed light curves, in two luminosity bins.  The
 right hand panels show Fourier power spectra for the same light
 curves, after subtracting a contribution for the underlying galaxy of
 absolute magnitude $M_{R} = -22.6$ in both luminosity bins.
 \label{fig:fig6}}
\end{figure*}

The two galaxies in Fig.~\ref{fig:fig4} were chosen so that the effect
is clear to the eye, but there is also a significant effect on the
statistical properties of variation when the correction is applied to a
large sample of light curves.  The right hand panel of 
Fig.~\ref{fig:fig5} shows the power spectrum for the sample of 302
light curves in the left hand panel, but with the contribution of the
underlying galaxy subtracted as described above.  It is clear that in
the right hand panel the two power spectra are indistinguishable within
the errors, implying that the residual variation is close to achromatic.
We can examine this model in a bit more detail by splitting the sample
into low and high luminosity objects to see whether a single luminosity
for the underlying galaxy really is sufficient.  In Fig.~\ref{fig:fig6}
we see power spectra as for Fig.~\ref{fig:fig5}, but in two luminosity
bins containing 125 and 177 light curves for low and high luminosity
quasars respectively.  In each case a galaxy luminosity of
$M_{R} = -22.6$ has been used, and it is clear that for the less
luminous quasars this works well.  For the more luminous quasars a
somewhat brighter underlying galaxy would be needed to produce
achromatic variation.

\subsection{Colour changes in accretion discs}

In most well studied systems which are believed to be powered by
accretion discs, variations in brightness are accompanied by colour
changes.  These are usually such that the object becomes bluer with
increasing luminosity.  Well known examples of this come from
dwarf novae \cite{h90} and Seyfert galaxies \cite{c91,p99}.
The development of accretion disc models has not so far reached the
point where quantitative predictions suitable for comparing with
observations have been made as to the nature of colour changes, so in
interpreting observed colour changes it is necessary to rely on
more general conceptual considerations.

The standard model of a geometrically thick, optically thick disc
\cite{s73} has a temperature gradient such that it becomes cooler and
redder with increasing radius.  It is not perhaps surprising that it
can be shown that the non-linear oscillations of such a disc can lead
to temperature and hence colour changes.  Radial temperature changes in
thermally unstable slim accretion discs have been examined in some
detail with the aid of numerical simulations by Honma et al. (1991).
They show that a hot wave propagates outwards as instabilities occur
in the inner disc, leading to a change in integrated colour, in the
sense that as the disc becomes brighter it becomes bluer.  This is
consistent with the observations of Seyfert galaxies \cite{c91,p99},
and also for the type of quasar variation seen in Fig.~\ref{fig:fig4}.
The achromatic variation seen in Fig.~\ref{fig:fig1} would not be
consistent with this model, and nor would the light curves shown in
Fig.~\ref{fig:fig2}.  Here, basically achromatic variation is
modified by the smoothing out in the red of the sharper features in
the blue light curves.

An alternative approach to modelling optical variability in accretion
discs \cite{m94} invokes a cellular automaton mechanism.  The accretion
disc achieves a self-organised critical state in which a $1/f$ spectrum
of fluctuations is produced.  This cellular automaton model appears to
be the most promising approach to modelling disc instabilities at the
moment.  It has been developed by Kawaguchi et al. (1998) to
investigate variability in AGN.  They use it to produce simulated light
curves with a power law spectrum of variations which can be compared
with observations \cite{h01}.  For Seyfert galaxies there is good
agreement, although for more luminous quasars the observed spectrum of
variations is flatter.  Although no quantitative predictions have so
far been made for colour changes, it is plausible that in a disc with a
strong temperature and hence colour gradient these avalanches will
produce chromatic effects as the flux varies.  The problem here is
that any variability would be on a very short timescale, and
inconsistent with the observations of optical variations.  Definitive
answers to these points must await model predictions for colour changes
in the different types of accretion disc.

\subsection{Colour changes from microlensing}

An alternative explanation for the effects shown in Fig.~\ref{fig:fig5}
comes from microlensing.  In this picture, the observed variation is
produced by the gravitational microlensing effect of a population of
planetary or sub-stellar mass bodies along the line of sight to the
quasar \cite{h96}.  Although the microlensing of a point source
produces a strictly achromatic light curve, for an accretion disc with
a radial temperature gradient this is not necessarily so.  For example,
if the nucleus is a compact blue source unresolved by the
characteristic mass of the microlenses, but in the red the source is
less compact and partially resolved by the lensing objects, colour
changes may be observed as the source is microlensed.  The effect of
this can be seen in the results of numerical simulation of microlensing
of unresolved and partially resolved sources \cite{l93,s87} where
the smaller power for the resolved sources is well illustrated.
The application of this model to the microlensing of an accretion
disc with a compact blue nucleus and a strong radial colour gradient
would be to produce less power in the red than the blue fluctuations,
providing that the emitting region of the disc is comparable with the
Einstein radius of the lenses.

These ideas have been put on a firmer and more quantitative footing
in a recent paper by Yonehara et al. (1999).  They carry out numerical
simulations of the microlensing of accretion disks by a compact body,
and show that for a standard optically thick accretion disc with a
radial temperature gradient colour changes will be seen, whereas for an
optically thin (advection dominated) disc where most of the emission
takes place at the inner edge, the variation will be achromatic.
Yonehara et al. (1999) base their modelling on the multiply lensed
quasar system Q2237+0305.  The individual images in this system have
been shown to vary achromatically when microlensed \cite{co91,hr94}
which would imply an optically thin disc.  However, to account for
the colour changes implied in Fig.~\ref{fig:fig5} and illustrated in
the light curves, an accretion disc of greater optical depth would be
required.

There is one other aspect of microlensing which has relevance to the
quasar light curves, especially those in Fig.~\ref{fig:fig2}.  In a
situation where the accretion disc has a blue compact core embedded in
redder outerparts, then lensing can produce cusp-like features from
the unresolved central region which are smoothed out in the resolved
red light source.  The idea can be seen from the synthetic light curves
for more and less resolved sources in Schneider \& Weiss (1987). 
This provides an explaination for the way sharp blue features become
smoother in the red in Fig.~\ref{fig:fig5}.

\section{Discussion}

It is clear from examination of Figs. 1-4 that colour changes can
occur in quasar light curves.  This result is consistent with
earlier work for Seyfert galaxies \cite{c91} and quasars \cite{c97},
but also shows that there are apparently different modes of colour 
change.  We tentatively identify the following patterns:

\begin{enumerate}

\item  Achromatic, no indication of any colour change.

\item  Light curves showing the same structure in red and blue, but
 with a scale change.

\item  The apparent bottoming out of red light curves as blue light
 continues to decline. 

\item  Colour change mainly characterised by the smoothing in the red
 of sharp features in the blue.

\end{enumerate}

It is quite possible that no single model will be found that can
account for such a diverse range of features in the light curves.
On the other hand, theories of quasar
variability should make predictions about colour changes, which
may well be similar and require detailed comparison with good data
to distinguish them.  Fig.~\ref{fig:fig5} implies that even
with a very simple model for the underlying galaxy, the mean power
spectra for red and blue light can be brought into coincidence.

Although it is clear that the presence of an underlying red galaxy
plays a significant role in AGN colour changes, it cannot account for
all the features seen in the two-colour light curves.  Examination
of Fig.~\ref{fig:fig2} shows that there are features in the
light curves which cannot be reconciled in this way.  To be more
specific, certain cusp-like features in the blue light curves appear
as if smoothed out in the red, although there is typically little
overall change in colour between maximum and minimum light.

At present, theoretical understanding of accretion disc instabilities
is not sufficiently well developed that such detailed features can be
modelled.  It is certainly possible that an event in the blue centre of
a nucleus could propagate into the redder outerparts and become blurred
in the process, but the apparent symmetry of the colour changes imposes
constraints on such a mechanism.  In disc instability models where the
emission is localised to a particular region of the disc, there may
well be colour changes associated with the outburst, but there is no
reason to believe that the structure of the event will be different
in red and blue passbands.

Microlensing does provide a possible explanation both for overall
changes in colour, and for the smearing out in the red of rapid
brightness changes in the blue.  This effect relies on an accretion
disc of comparable size to the Einstein radii of the lenses, but
given this it is a prediction of microlensing variations that if
colour changes are observed at all they will be such that features in
the blue will be smoothed out in the red.  Furthermore, the larger
the colour change, the more pronounced this smoothing will be.

\section{Conclusions}

In this paper we have investigated colour changes in a large sample of
quasars, monitored yearly in the $B_{J}$ and $R$ bands over a period
of 21 years.  Examination of the light curves suggests that the
variations involve colour changes for many quasars, and Fourier
power spectrum analysis for a 15 year homogeneous subset of the
data confirms that there is more power on all
timescales in the blue passband compared with red.  This effect may at
least in part be due to the contribution of red light from the
underlying host galaxy, and can be corrected for in such a way that
the variation comes close to being achromatic for all quasars.

Current theories of accretion disc instability can explain colour
changes seen in observations of Seyfert galaxies.  The standard
optically thick accretion disk is characterised by a radial temperature
gradient which when perturbed will tend to exhibit changes in
integrated colour.  However, they do not appear to be able to
produce the smearing out in the red of blue features which is a
characteristic of many quasar light curves.

An alternative explanation for the observed variability of quasars is
that it is caused by the microlensing effects of a population of
planetary or sub-stellar mass bodies distributed along the line of
sight.  In this case achromatic variation can be accounted for by
microlensing of an optically thin (advection dominated) disc.
Chromatic variations are possible with a conventional optically thick
disc with a radial colour gradient, which is large compared with the
Einstein radius of the lenses.  This can also explain the the observed
smearing out in the red of cusp-like features in the blue.  

\section*{Acknowledgements}

I wish to thank Alan Heavens for suggestions with regard to the
subtraction of the flux of the underlying host galaxy.

\newpage
\begin{table}
\section*{Appendix}
\caption{Details of photographic observations for the two-colour
 monitoring programme}
\begin{tabular}{llllr}
Plate & Date & Emulsion & Filter & Exposure \\
 & & & & \\
 R 6196 & 80-08-04 & IIIaF & RG 630 &  90.0 \\
 R 6200 & 80-08-05 & IIIaF & RG 630 &  90.0 \\
 R 6209 & 80-08-06 & IIIaF & RG 630 &  90.0 \\
 J 6229 & 80-08-09 & IIIaJ & GG 395 &  65.0 \\
 R 6230 & 80-08-09 & IIIaF & RG 630 &  90.0 \\
 R 6238 & 80-08-10 & IIIaF & RG 630 &  90.0 \\
 J 7099 & 81-08-01 & IIIaJ & GG 395 &  75.0 \\
 R 7259 & 81-10-06 & IIIaF & RG 630 &  15.0 \\
 R 8005 & 82-08-14 & IIIaF & RG 630 &  90.0 \\
 J 8008 & 82-08-16 & IIIaJ & GG 395 &  60.0 \\
 R 8649 & 83-07-09 & IIIaF & RG 630 &  90.0 \\
 R 8671 & 83-07-14 & IIIaF & RG 630 &  90.0 \\
 R 8717 & 83-08-03 & IIIaF & RG 630 &  90.0 \\
 R 8723 & 83-08-04 & IIIaF & RG 630 &  90.0 \\
 J 8727 & 83-08-05 & IIIaJ & GG 395 &  65.0 \\
 J 8728 & 83-08-05 & IIIaJ & GG 395 &  65.0 \\
 J 8729 & 83-08-05 & IIIaJ & GG 395 &  65.0 \\
 J 8730 & 83-08-05 & IIIaJ & GG 395 &  65.0 \\
 J 9322 & 84-05-26 & IIIaJ & GG 395 &  60.0 \\
 R 9323 & 84-05-26 & IIIaF & RG 630 &  80.0 \\
 J 9327 & 84-05-27 & IIIaJ & GG 395 &  60.0 \\
 J 9335 & 84-05-28 & IIIaJ & GG 395 &  60.0 \\
 R 9336 & 84-05-28 & IIIaF & RG 630 &  90.0 \\
 J 9338 & 84-05-29 & IIIaJ & GG 395 &  60.0 \\
 R 9365 & 84-06-03 & IIIaF & RG 630 &  90.0 \\
 R 9374 & 84-06-04 & IIIaF & RG 630 &  90.0 \\
 J10260 & 85-06-14 & IIIaJ & GG 395 &  70.0 \\
 J10264 & 85-06-15 & IIIaJ & GG 395 &  70.0 \\
 J10271 & 85-06-16 & IIIaJ & GG 395 &  70.0 \\
 J11344 & 86-09-04 & IIIaJ & GG 395 &  80.0 \\
 R11345 & 86-09-05 & IIIaF & RG 630 &  90.0 \\
 R11351 & 86-09-06 & IIIaF & RG 630 &  80.0 \\
 R11359 & 86-09-07 & IIIaF & RG 630 &  75.0 \\
 R11360 & 86-09-07 & IIIaF & RG 630 &  75.0 \\
 J11387 & 86-09-25 & IIIaJ & GG 395 &  90.0 \\
 R11919 & 87-05-25 & IIIaF & RG 630 & 128.9 \\
 J11949 & 87-06-04 & IIIaJ & GG 395 &  75.0 \\
 J11971 & 87-06-22 & IIIaJ & GG 395 & 104.4 \\
 J12012 & 87-07-18 & IIIaJ & GG 395 &  75.0 \\
 J12023 & 87-07-22 & IIIaJ & GG 395 &  28.5 \\
 R12084 & 87-08-21 & IIIaF & RG 630 & 130.0 \\
 R12136 & 87-09-14 & IIIaF & RG 630 & 130.0 \\
 J12602 & 88-06-12 & IIIaJ & GG 395 & 120.0 \\
 R12626 & 88-06-19 & IIIaF & RG 630 & 106.1 \\
 J12648 & 88-07-12 & IIIaJ & GG 395 & 100.0 \\
 J12656 & 88-07-14 & IIIaJ & GG 395 & 100.0 \\
 J12657 & 88-07-14 & IIIaJ & GG 395 & 100.0 \\
 R12669 & 88-07-18 & IIIaF & RG 630 & 150.0 \\
 R12740 & 88-09-04 & IIIaF & RG 630 & 140.0 \\
 R12771 & 88-10-07 & IIIaF & RG 630 & 140.0 \\
 R13207 & 89-07-27 & IIIaF & RG 630 &  90.0 \\
 R13251 & 89-08-26 & IIIaF & RG 630 & 120.0 \\
 J13252 & 89-08-26 & IIIaJ & GG 395 &  60.0 \\
 J13269 & 89-09-03 & IIIaJ & GG 395 &  65.0 \\
 R13270 & 89-09-03 & IIIaF & RG 630 &  75.0 \\
 J13315 & 89-09-24 & IIIaJ & GG 395 &  60.0 \\
 J13316 & 89-09-24 & IIIaJ & GG 395 &  60.0 \\
 R13338 & 89-09-28 & IIIaF & RG 630 & 100.0 \\
 J13760 & 90-06-23 & IIIaJ & GG 395 &  60.0 \\
\end{tabular}
\end{table}

\begin{table}
\vspace{10mm}
\begin{tabular}{llllr}
Plate & Date & Emulsion & Filter & Exposure \\
 & & & & \\
 R13767 & 90-06-29 & IIIaF & RG 630 &  90.0 \\
 J13789 & 90-08-19 & IIIaJ & GG 395 &  60.0 \\
 R13790 & 90-08-19 & IIIaF & RG 630 &  84.2 \\
 J13819 & 90-09-10 & IIIaJ & GG 395 &  60.0 \\
 R13830 & 90-09-16 & IIIaF & RG 630 &  95.0 \\
 J13837 & 90-09-17 & IIIaJ & GG 395 &  65.0 \\
 R13909 & 90-10-12 & IIIaF & RG 630 & 110.0 \\
 R14396 & 91-07-07 & IIIaF & RG 630 &  90.0 \\
 R14417 & 91-07-19 & IIIaF & RG 630 &  90.0 \\
 R14422 & 91-07-20 & IIIaF & RG 630 & 100.0 \\
 J14449 & 91-08-07 & IIIaJ & GG 395 &  60.0 \\
 R14502 & 91-09-01 & IIIaF & RG 630 & 100.0 \\
 J14640 & 91-10-31 & IIIaJ & GG 395 &  60.0 \\
 J14643 & 91-11-01 & IIIaJ & GG 395 &  80.0 \\
 J14648 & 91-11-07 & IIIaJ & GG 395 &  60.0 \\
 R15005 & 92-06-03 & IIIaF & RG 630 &  90.0 \\
 J15006 & 92-06-03 & IIIaJ & GG 395 &  80.0 \\
 J15034 & 92-07-01 & IIIaJ & GG 395 & 100.0 \\
 R15045 & 92-07-03 & IIIaF & RG 630 &  62.1 \\
 R15073 & 92-07-26 & IIIaF & RG 630 &  90.0 \\
 R15086 & 92-07-28 & IIIaF & RG 630 &  90.0 \\
 J15102 & 92-08-01 & IIIaJ & GG 395 &  80.0 \\
 J15563 & 93-05-24 & IIIaJ & GG 395 &  65.0 \\
 R15564 & 93-05-24 & IIIaF & RG 630 &  85.0 \\
 R15575 & 93-05-26 & IIIaF & RG 630 &  90.0 \\
 R15673 & 93-08-09 & IIIaF & RG 630 &  90.0 \\
 R15684 & 93-08-11 & IIIaF & RG 630 &  90.0 \\
 J15795 & 93-10-05 & IIIaJ & GG 395 &  60.0 \\
 J15796 & 93-10-06 & IIIaJ & GG 395 & 100.0 \\
 J15801 & 93-10-07 & IIIaJ & GG 395 &  90.0 \\
 R16149 & 94-06-15 & IIIaF & RG 630 &  90.0 \\
 R16160 & 94-06-17 & IIIaF & RG 630 &  66.8 \\
 J16177 & 94-07-06 & IIIaJ & GG 395 & 110.0 \\
 J16180 & 94-07-07 & IIIaJ & GG 395 & 110.0 \\
 J16183 & 94-07-08 & IIIaJ & GG 395 & 110.0 \\
 R16208 & 94-07-29 & IIIaF & RG 630 &  70.0 \\
 R16224 & 94-08-09 & IIIaF & RG 630 & 135.0 \\
 J16230 & 94-08-10 & IIIaJ & GG 395 & 110.0 \\
 R16678 & 95-07-06 & 4415  & RG 630 &  90.0 \\
 J16700 & 95-07-29 & IIIaJ & GG 395 &  60.0 \\
 R16703 & 95-07-30 & 4415  & RG 630 &  90.0 \\
 J16728 & 95-08-20 & IIIaJ & GG 395 &  60.0 \\
 J16731 & 95-08-23 & IIIaJ & GG 395 &  60.0 \\
 J16741 & 95-08-25 & IIIaJ & GG 395 &  70.0 \\
 R16747 & 95-08-26 & 4415  & RG 630 &  90.0 \\
 R16754 & 95-08-28 & 4415  & RG 630 &  90.0 \\
 R17108 & 96-05-23 & 4415  & RG 630 &  90.0 \\
 J17117 & 96-05-25 & IIIaJ & GG 395 &  90.0 \\
 J17167 & 96-08-07 & IIIaJ & GG 395 &  60.0 \\
 J17173 & 96-08-11 & IIIaJ & GG 395 &  60.0 \\
 J17189 & 96-08-13 & IIIaJ & GG 395 &  60.0 \\
 R17197 & 96-08-14 & 4415  & RG 630 & 105.0 \\
 R17232 & 96-09-11 & 4415  & RG 630 & 120.0 \\
 R17254 & 96-09-14 & 4415  & RG 630 & 110.0 \\
 R17592 & 97-06-07 & 4415  & RG 630 &  90.0 \\
 R17599 & 97-06-08 & 4415  & RG 630 & 120.0 \\
 R17611 & 97-06-10 & 4415  & RG 630 & 120.0 \\
 R17615 & 97-06-11 & 4415  & RG 630 & 120.0 \\
 J17774 & 97-10-19 & IIIaJ & GG 395 &  60.0 \\
 J10292 & 85-06-24 & IIIaJ & GG 395 &  90.0 \\
 R10302 & 85-06-26 & IIIaF & RG 630 & 100.0 \\
 R10306 & 85-06-27 & IIIaF & RG 630 & 105.0 \\
 R10341 & 85-07-20 & IIIaF & RG 630 & 105.0 \\
\end{tabular}
\end{table}

\begin{table}
\vspace{10mm}
\begin{tabular}{llllr}
Plate & Date & Emulsion & Filter & Exposure \\
 & & & & \\
 R10349 & 85-07-23 & IIIaF & RG 630 &  90.0 \\
 J11319 & 86-08-29 & IIIaJ & GG 395 &  80.0 \\
 J11334 & 86-09-02 & IIIaJ & GG 395 &  85.0 \\
 J17775 & 97-10-21 & IIIaJ & GG 395 &  60.0 \\
 J17779 & 97-10-22 & IIIaJ & GG 395 &  60.0 \\
 J17788 & 97-10-24 & IIIaJ & GG 395 &  60.0 \\
 R18028 & 98-05-28 & 4415  & RG 630 &  90.0 \\
 J18110 & 98-09-17 & IIIaJ & GG 395 &  60.0 \\
 R18484 & 99-07-09 & 4415  & RG 630 &  90.0 \\
 J18492 & 99-07-19 & IIIaJ & GG 395 &  35.9 \\
 J18503 & 99-08-03 & IIIaJ & GG 395 &  60.0 \\
 J18514 & 99-08-05 & IIIaJ & GG 395 &  60.0 \\
 R18521 & 99-08-06 & 4415  & RG 630 & 110.0 \\
 J18522 & 99-08-06 & IIIaJ & GG 395 &  49.6 \\
 R18524 & 99-08-09 & 4415  & RG 630 &  90.0 \\
 R18529 & 99-08-10 & 4415  & RG 630 &  90.0 \\
 R18853 & 00-06-02 & 4415  & RG 630 &  90.0 \\
 R18855 & 00-06-03 & 4415  & RG 630 &  90.0 \\
 R18876 & 00-07-01 & 4415  & RG 630 &  90.0 \\
 R18882 & 00-07-06 & 4415  & RG 630 &  90.0 \\
 J18919 & 00-08-06 & IIIaJ & GG 395 &  60.0 \\
 J18926 & 00-08-29 & IIIaJ & GG 395 &  60.0 \\
 J18945 & 00-09-18 & IIIaJ & GG 395 &  60.0 \\
 J18947 & 00-09-19 & IIIaJ & GG 395 &  60.0 \\
 R19209 & 01-06-20 & 4415  & RG 630 &  90.0 \\
 R19210 & 01-06-20 & 4415  & RG 630 &  90.0 \\
 R19217 & 01-06-25 & 4415  & RG 630 &  90.0 \\
 R19219 & 01-06-27 & 4415  & RG 630 &  90.0 \\
 J19220 & 01-06-27 & IIIaJ & GG 395 &  60.0 \\
 J19225 & 01-06-29 & IIIaJ & GG 395 &  60.0 \\
 J19235 & 01-07-22 & IIIaJ & GG 395 &  49.0 \\
 J19236 & 01-07-22 & IIIaJ & GG 395 &  60.0 \\
\end{tabular}
\end{table}

\end{document}